\documentstyle[twoside]{article}

\catcode`\@=11
\long\def\@makefntext#1{
\protect\noindent \hbox to 3.2pt {\hskip-.9pt  
$^{{\eightrm\@thefnmark}}$\hfil}#1\hfill}		

\def\thefootnote{\fnsymbol{footnote}}
\def\@makefnmark{\hbox to 0pt{$^{\@thefnmark}$\hss}}	
	
\def\ps@myheadings{\let\@mkboth\@gobbletwo
\def\@oddhead{\hbox{}
\rightmark\hfil\eightrm\thepage}   
\def\@oddfoot{}\def\@evenhead{\eightrm\thepage\hfil
\leftmark\hbox{}}\def\@evenfoot{}
\def\sectionmark##1{}\def\subsectionmark##1{}}

\oddsidemargin=\evensidemargin
\renewcommand{\thefootnote}{\fnsymbol{footnote}}

\newcounter{sectionc}\newcounter{subsectionc}\newcounter{subsubsectionc}
\renewcommand{\section}[1] {\vspace{12pt}\addtocounter{sectionc}{1} 
\setcounter{subsectionc}{0}\setcounter{subsubsectionc}{0}\noindent 
	{\tenbf\thesectionc. #1}\par\vspace{5pt}}
\renewcommand{\subsection}[1] {\vspace{12pt}\addtocounter{subsectionc}{1} 
	\setcounter{subsubsectionc}{0}\noindent 
	{\bf\thesectionc.\thesubsectionc. {\kern1pt \bfit #1}}\par\vspace{5pt}}
\renewcommand{\subsubsection}[1] {\vspace{12pt}\addtocounter{subsubsectionc}{1}
	\noindent{\tenrm\thesectionc.\thesubsectionc.\thesubsubsectionc.
	{\kern1pt \tenit #1}}\par\vspace{5pt}}
\newcommand{\nonumsection}[1] {\vspace{12pt}\noindent{\tenbf #1}
	\par\vspace{5pt}}

\newcounter{appendixc}
\newcounter{subappendixc}[appendixc]
\newcounter{subsubappendixc}[subappendixc]
\renewcommand{\thesubappendixc}{\Alph{appendixc}.\arabic{subappendixc}}
\renewcommand{\thesubsubappendixc}
	{\Alph{appendixc}.\arabic{subappendixc}.\arabic{subsubappendixc}}

\renewcommand{\appendix}[1] {\vspace{12pt}
        \refstepcounter{appendixc}
        \setcounter{figure}{0}
        \setcounter{table}{0}
        \setcounter{lemma}{0}
        \setcounter{theorem}{0}
        \setcounter{corollary}{0}
        \setcounter{definition}{0}
        \setcounter{equation}{0}
        \renewcommand{\thefigure}{\Alph{appendixc}.\arabic{figure}}
        \renewcommand{\thetable}{\Alph{appendixc}.\arabic{table}}
        \renewcommand{\theappendixc}{\Alph{appendixc}}
        \renewcommand{\thelemma}{\Alph{appendixc}.\arabic{lemma}}
        \renewcommand{\thetheorem}{\Alph{appendixc}.\arabic{theorem}}
        \renewcommand{\thedefinition}{\Alph{appendixc}.\arabic{definition}}
        \renewcommand{\thecorollary}{\Alph{appendixc}.\arabic{corollary}}
        \renewcommand{\theequation}{\Alph{appendixc}.\arabic{equation}}
        \noindent{\tenbf Appendix \theappendixc #1}\par\vspace{5pt}}
\newcommand{\subappendix}[1] {\vspace{12pt}
        \refstepcounter{subappendixc}
        \noindent{\bf Appendix \thesubappendixc. {\kern1pt \bfit #1}}
	\par\vspace{5pt}}
\newcommand{\subsubappendix}[1] {\vspace{12pt}
        \refstepcounter{subsubappendixc}
        \noindent{\rm Appendix \thesubsubappendixc. {\kern1pt \tenit #1}}
	\par\vspace{5pt}}

\newcommand{\textlineskip}{\baselineskip=13pt}
\newcommand{\smalllineskip}{\baselineskip=10pt}

\def\eightcirc{
\begin{picture}(0,0)
\put(4.4,1.8){\circle{6.5}}
\end{picture}}
\def\eightcopyright{\eightcirc\kern2.7pt\hbox{\eightrm c}} 

\def\abstracts#1#2#3{{
	\centering{\begin{minipage}{4.5in}\baselineskip=10pt\footnotesize
	\parindent=0pt #1\par 
	\parindent=15pt #2\par
	\parindent=15pt #3
	\end{minipage}}\par}} 

\newcommand{\bibit}{\nineit}

\renewenvironment{thebibliography}[1]
	{\frenchspacing
	 \ninerm\baselineskip=11pt
	 \begin{list}{\arabic{enumi}.}
	{\usecounter{enumi}\setlength{\parsep}{0pt}
	 \setlength{\leftmargin 12.7pt}{\rightmargin 0pt} 
	 \setlength{\itemsep}{0pt} \settowidth
	{\labelwidth}{#1.}\sloppy}}{\end{list}}

\newcounter{itemlistc}
\newcounter{romanlistc}
\newcounter{alphlistc}
\newcounter{arabiclistc}

\newcommand{\fcaption}[1]{
        \refstepcounter{figure}
        \setbox\@tempboxa = \hbox{\footnotesize Fig.~\thefigure. #1}
        \ifdim \wd\@tempboxa > 5in
           {\begin{center}
        \parbox{5in}{\footnotesize\smalllineskip Fig.~\thefigure. #1}
            \end{center}}
        \else
             {\begin{center}
             {\footnotesize Fig.~\thefigure. #1}
              \end{center}}
        \fi}

\newcommand{\tcaption}[1]{
        \refstepcounter{table}
        \setbox\@tempboxa = \hbox{\footnotesize Table~\thetable. #1}
        \ifdim \wd\@tempboxa > 5in
           {\begin{center}
        \parbox{5in}{\footnotesize\smalllineskip Table~\thetable. #1}
            \end{center}}
        \else
             {\begin{center}
             {\footnotesize Table~\thetable. #1}
              \end{center}}
        \fi}

\def\@citex[#1]#2{\if@filesw\immediate\write\@auxout
	{\string\citation{#2}}\fi
\def\@citea{}\@cite{\@for\@citeb:=#2\do
	{\@citea\def\@citea{,}\@ifundefined
	{b@\@citeb}{{\bf ?}\@warning
	{Citation `\@citeb' on page \thepage \space undefined}}
	{\csname b@\@citeb\endcsname}}}{#1}}

\newif\if@cghi
\def\cite{\@cghitrue\@ifnextchar [{\@tempswatrue
	\@citex}{\@tempswafalse\@citex[]}}
\def\citelow{\@cghifalse\@ifnextchar [{\@tempswatrue
	\@citex}{\@tempswafalse\@citex[]}}
\def\@cite#1#2{{$\null^{#1}$\if@tempswa\typeout
	{IJCGA warning: optional citation argument 
	ignored: `#2'} \fi}}

\def\pmb#1{\setbox0=\hbox{#1}
	\kern-.025em\copy0\kern-\wd0
	\kern.05em\copy0\kern-\wd0
	\kern-.025em\raise.0433em\box0}

\def\fnt#1#2{\footnotetext{\kern-.3em
	{$^{\mbox{\scriptsize #1}}$}{#2}}}

\def\fpage#1{\begingroup
\voffset=.3in
\thispagestyle{empty}\begin{table}[b]\centerline{\footnotesize #1}
	\end{table}\endgroup}

\def\runninghead#1#2{\pagestyle{myheadings}
\markboth{{\protect\footnotesize\it{\quad #1}}\hfill}
{\hfill{\protect\footnotesize\it{#2\quad}}}}
\font\tenrm=cmr10
\font\tenit=cmti10 
\font\tenbf=cmbx10
\font\bfit=cmbxti10 at 10pt
\font\ninerm=cmr9
\font\nineit=cmti9

\font\eightrm=cmr8

\def\qed{\hbox{${\vcenter{\vbox{			
   \hrule height 0.4pt\hbox{\vrule width 0.4pt height 6pt
   \kern5pt\vrule width 0.4pt}\hrule height 0.4pt}}}$}}

\begin{document}
\runninghead{Vertex Operators for AdS3 with Ramond Background}
{Vertex Operators for AdS3 with Ramond Background}

\normalsize\textlineskip
\thispagestyle{empty}
\setcounter{page}{1}


\vspace*{0.88truein}

\fpage{1}
\centerline{\bf VERTEX OPERATORS FOR ADS3}
\vspace*{0.035truein}
\centerline{\bf WITH RAMOND BACKGROUND}
\vspace*{0.37truein}
\centerline{\footnotesize L. DOLAN}
\vspace*{0.015truein}
\centerline{\footnotesize\it Department of Physics and Astronomy, University
of North Carolina} 
\baselineskip=10pt
\centerline{\footnotesize\it Chapel HIll, NC 27599-3255, USA}
\vspace*{0.225truein}

\vspace*{0.21truein}
\abstracts{This review gives results on vertex 
operators for the 
Type IIB superstring in an AdS3 x S3 background with Ramond-Ramond flux,
which were presented at Strings 2000.
Constraint equations for these vertex operators are derived,
and their components are 
shown to satisfy the supergravity linearized equations of motion for the 
six-dimensional (2,0) theory of a supergravity and tensor
multiplet expanded around AdS3 x S3 spacetime.}{}{}


\vspace*{1pt}\textlineskip	
\section{Introduction}	
\vspace*{-0.5pt}
\def\half{{1\over2}}
\noindent
The conjectured duality between M-theory or Type IIB string theory
on anti-de Sitter (AdS) space and the conformal field theory on
the boundary of AdS space may be useful in giving a controlled
systematic approximation for strongly coupled gauge theories. 
Examples with maximal supersymmetry correspond to a set of $p$-branes
whose near horizon geometry
looks like $AdS_{p+2}\times S^{{\cal D}-p-2}$, where ${\cal D} =$ 10 or 11
for branes in string or M-theory, and $p=2,3,5$.
The formulation of vertex operators and
string theory tree amplitudes for the IIB superstring on $AdS_5\times S^5$ 
will allow access
to the dual conformal $SU(N)$ gauge field theory $CFT_4$ at large $N$, but
{\it small} fixed `t Hooft coupling $x = g^2_{YM} N$ in the dual
picture, as
$(g^2_{YM} N)^\half (4\pi)^\half = R^2_{sph}/{\alpha'}$.
Presently only the large $N$, and {\it large} fixed `t Hooft coupling $x$
limit is accessible in the $CFT$, since only the supergravity limit
($\alpha'\rightarrow 0$) of the correlation functions 
of the AdS theory is known. 

In this talk we discuss a case with non-maximal supersymmetry that is
related to a system with a $D1$-brane and a $D5$-brane.
Its quantizable worldsheet action$^1$ describes the IIB string on 
$AdS_3\times S^3\times M$ with background Ramond-Ramond flux, where $M$
is $T^4$ or $K3$. The vertex operators for this model can be computed explicitly
in the bulk.$^2$ Correlation functions constructed from these
vertex operators, restricted to the boundary of $AdS_3$, would be those for a
two-dimensional space-time conformal field theory.
We work to leading order in $\alpha'$, but because of the
high degree of symmetry of the model,
we expect our result for the vertex operators to
be exact. Tree level $n$-point correlation functions for $n\ge 4$
presumably have $\alpha'$ corrections, since the worldsheet theory is
not a free conformal field theory.  But 
\textheight=7.8truein
\setcounter{footnote}{0}
\renewcommand{\thefootnote}{\alph{footnote}}
\noindent 
there may be sufficiently many
symmetry currents to determine the tree level 
correlation functions exactly in $\alpha'$ as well.$^3$   

In terms of Berkovits-Vafa-Witten worldsheet variables,  
constraint equations for the vertex operators in the flat space ${\cal R}^6$
follow from
the physical state conditions coming from an $N=4$ superconformal algebra.
We generalize$^2$ these constraint equations to 
$AdS_3\times S^3$ for the vertex operators
of the massless states that are independent of the compactification $M$.
We then solve the constraints and identify the components
of the vertex operators as supergravity fields, that satisfy
the $D=6$, $N=(2,0)$ theory$^4$ linearized around the $AdS_3\times S^3$
background.  
 
Recent work$^{5,6}$ discusses covariant ten-dimensional worldsheet
variables and extends our analysis to vertex operators on $AdS_5$ in a
spinor formulation.

\def\sqr#1#2{{\vbox{\hrule height.#2pt\hbox{\vrule width
.#2pt height#1pt \kern#1pt\vrule width.#2pt}\hrule height.#2pt}}}
\def\Box{\mathchoice\sqr64\sqr64\sqr{4.2}3\sqr33}  
\def\e{\epsilon}
\def\N{{\nabla}}
\def\Nb{{\bar\nabla}}
\def\p{{\partial}}
\def\pb{{\bar\partial}} 
\def\Tr{{\rm Tr}}
\def\apm{\alpha'}
\def\zbar{\bar z}
\def\qbar{\bar q}
\def\half{{\textstyle{1\over 2}}}
\def\hhalf{{scriptstyle{1\over 2}}}
\def\quar{{\textstyle{1\over 4}}}
\def\third{{\textstyle{1\over3}}}
\def\tthird{{\textstyle{2\over3}}}
\def\summ{\sum_{n = -\infty}^{\infty}}
\def\ooint{{1 \over {2 \pi i}} \oint}
\def\tr{\hbox{tr}_\alpha}
\def\re{{\bf R}}
\def\ze{{\bf Z}}
\def\ce{{\bf C}}
\def\ag{{g}}
\def\ah{{h}}
\def\dt{{\cdot}}
\def\nox{{\scriptstyle{\times \atop \times}}}
\def\zg{\zeta}
\def\ref#1{$^{[#1]}$}
\def\zd{{(z - \zeta)}}
\def\ag{\alpha}
\def\bg{\beta}
\def\gg{\gamma}
\def\emp{\hbox{\O}}
\def\sH{{\cal H}}
\def\hf{h^\mu_r}
\def\ih{{\hat \imath}}
\def\jh{{\hat \jmath}}
\def\thr{{\bf 3}}
\def\one{{\bf 1}}
\def\two{{\bf 2}}
\def\vp#1{\hbox{\vbox{#1\vskip5pt}}}
\def\Thf #1. #2.{\Theta\left[{\textstyle{#1\atop #2}}\right]}
\def\Chf #1. #2.{\bar\Theta\left[{\textstyle{#1\atop #2}}\right]}
\def\Gg{\Gamma}
\def\star{\mathop{\ast}}
\def\noblackbox{\overfullrule=0pt}
\def\t{\theta}
\def\s{\sigma}
\noblackbox
\section{Formulating Strings on $AdS$}
\noindent
In the Ramond-Neveu-Schwarz (RNS) formalism,
the worldsheet action for strings on $AdS$ space with background
Ramond-Ramond flux involves 2d spin fields. These violate 
superconformal worldsheet symmetry, since the worldsheet
supercurrents are not local with respect to the spin fields,
and their presence makes the worldsheet theory difficult to understand.

For the Type IIB superstring on 
$AdS_3\times S^3$ case, a sigma model$^1$ with conventional local interactions 
(no spin fields in the action) was found using the supergroup
$PSU(2|2)$ as target, coupled to ghost fields $\rho$ and $\sigma$.
The spacetime symmetry group is $PSU(2|2)\times PSU(2|2)$,
acting by left and right multiplication on the group manifold,
{\it i.e.}  by $g\to agb^{-1}$ where $g$ is a $PSU(2|2)$-valued field, 
and $a,b\in PSU(2|2)$ are the symmetry group's Lie algebra elements.
The supergroup is generated by the super Lie algebra with 12 bosonic
generators forming a subalgebra $SO(4)\times SO(4)$ 
together with 16 odd generators.
Hence our model has non-maximal supersymmetry with 16 supercharges.

The worldsheet field content generalizes 
the Berkovits-Vafa formalism which provides a manifest Lorentz covariant and
supersymmetric quantization on $R^6$. Its 
six bosonic fields $x^p(z,\bar z)$ contain
both left- and right-moving modes. In addition there are
left-moving fermi fields $ \t^a_L(z),
p^a_L(z)$ of spins 0 and 1, together with ghosts
$\s_L(z),\rho_L(z)$, and right-moving counterparts of all
these left-moving fields. These variables allow Ramond-Ramond background fields 
to be incorporated without adding spin fields to the worldsheet action as 
follows:
in the $AdS_3 \times S^3$ case, i.e. after adding RR background fields to the
worldsheet action, one can integrate out the  $p$'s, so that the model
has ordinary conformal fields  $x^p$, $\t^a$, $\bar\t^a$ (all now with
both left- and right-moving components) as well as the ghosts.
The $PSU(2|2)$-valued field $g$ is given in terms of $x,\t$, and $\bar\t$,
which are identified as coordinates on the supergroup manifold. 
In addition, Type IIB on $AdS_3 \times S^3\times M$ has worldsheet
variables describing the compactification on the 
four-dimensional space $M$. Their Virasoro currents have central charge $c=6$,
and will be labeled with a subscript $C$.  
 
\section{$N = 4$ Super Virasoro Generators }
\indent
The holomorphic $N=4$ superconformal generators with $c=6$ are 
given for flat space by$^1$
\begin{eqnarray}
T&=&-\half\partial x^m\partial x_m - p_a \partial\theta^a
-\half\partial\rho\partial\rho - \half\partial\sigma\partial\sigma
+ \partial^2 (\rho + i\sigma) + T_C\nonumber\\
G^+&=&-e^{-2\rho-i\sigma} (p)^4 + {\textstyle{i\over 2}}e^{-\rho} p_a p_b
\partial x^{ab}\nonumber\\ 
&\hskip 5pt & 
+e^{i\sigma} (-\half\partial x^m\partial x_m - p_a \partial\theta^a 
-\half\partial (\rho + i\sigma) \partial (\rho + i\sigma) 
+\half\partial^2  (\rho + i\sigma) ) \,+ G^+_C\nonumber\\
G^-&=&e^{-i\sigma} + G_C^- \nonumber\\
J&=&\partial (\rho + i\sigma) + J_C\nonumber\\
\tilde G^+&=& e^{iH_C} \,[ -e^{-3\rho-2i\sigma} (p)^4 + {\textstyle{i\over 2}}
e^{-2\rho-i\sigma} p_a p_b \partial x^{ab}\nonumber\\
&\hskip 7pt &
+ e^{-\rho} (-\half\partial x^m\partial x_m - p_a \partial\theta^a 
-\half\partial (\rho + i\sigma) \partial (\rho + i\sigma)\nonumber\\
&\hskip 7pt &
+\half\partial^2  (\rho + i\sigma) ] \,+ e^{-\rho-i\sigma} \tilde G_C^-
\nonumber\\
J^+&=& e^{\rho + i\sigma} J_C^+\nonumber\\
J^-&=& e^{-\rho - i\sigma} J_C^-\,.\nonumber\\
\end{eqnarray}
These currents are given in terms of the left-moving bosons
$\partial x^m, \rho, \sigma$, and the left-moving fermionic worldsheet 
fields $p^a,\t^a$, 
where $1\le m\le 6, 1\le a\le 4$.
There are corresponding anti-holomorphic expressions. 
Both sets of generators
are used to implement the physical state conditions on the vertex 
operators, a procedure which results in a set of string constraint equations.

\section{String Constraint Equations for the Vertex Operators}
\noindent
The expansion of the massless vertex operator in terms of the 
worldsheet fields is
\begin{equation}
V = \sum_{m,n = -\infty}^\infty\,
e^{m(i\s + \rho) + n(i\bar\s + \bar\rho)}\,
V_{m,n} (x, \t, \bar\t)\,.\end{equation}
In flat space, the constraints from the left and right-moving
worldsheet super Virasoro algebras are:
\begin{eqnarray}
(\N)^4 V_{1,n} &=& \N_a \,\N_b
\p^{a b} V_{1,n} = 0\nonumber\\
{\textstyle{1\over 6}}\e^{a b c d} \,\N_b \,\N_c \,\N_d
V_{1,n} &=& - i \N_b\, \p^{a b} V_{0,n}\nonumber\\
\N_a\,\N_b\, V_{0,n} - {\textstyle{i\over 2}} \e_{a b c d} \, \p^{cd}\,
V_{-1,n} &=& 0\,,\qquad \N_a \, V_{-1,n} = 0\,;\\
\bar\N^4 V_{n,1} &=& \bar\N_{\bar a}\bar\N_{\bar b}
\bar\p^{\bar a\bar b} V_{n,1} = 0\nonumber\\
{\textstyle{1\over 6}} \e^{\bar a\bar b\bar c\bar d}\bar
\N_{\bar b}\bar \N_{\bar c} \bar \N_{\bar d}
V_{n,1} &=& -i \bar \N_{\bar b} \bar\p^{\bar a\bar b} V_{n,0}\nonumber\\
\bar\N_{\bar a}\bar \N_{\bar b} V_{n,0}
- {\textstyle{i\over 2}} \bar\e_{\bar a \bar b \bar c \bar d}
\, \bar\p^{\bar c\bar d}\, V_{n,-1} &=& 0\,,
\qquad \bar\N_{\bar a} \, V_{n,-1} = 0
\end{eqnarray}
\begin{equation}\p^p\p_p V_{m,n} = 0
\end{equation}
for $-1\le m,n\le 1$, with the notation
$\N_a = d/ d\t^a$, $\bar\N_{\bar a} = d/ d\bar\t^{\bar a}$,
$\partial^{ab} = -\sigma^{p ab}\,\p_p$.
These equations were derived in flat space by requiring the
vertex operators to satisfy the physical state conditions
\begin{eqnarray}
G^-_0 V=\tilde G^-_0 V=\bar G^-_0 V=
\bar{\tilde G}^-_0 V =T_0 V=\bar T_0 V= 0,\nonumber\\
J_0 V = \bar J_0 V = 0\,,\quad G^+_0\tilde G^+_0 V=
\bar G^+_0\bar{\tilde G}^+_0 V=0\nonumber\\
\end{eqnarray}
where $T_n,\, G^\pm_n,\, \tilde  G^\pm_n,\,J_n, J^\pm_n$ and
corresponding barred generators
are the left and right $N=4$ worldsheet superconformal generators.
These conditions further imply
$V_{m,n} = 0$ for $m>1$ or $n>1$ or $m<1$ or $n<1$, leaving
nine non-zero components.

In $AdS_3\times S^3$ space, we generalize these equations as follows$^2$:
\begin{eqnarray}
F^4 V_{1,n} &=& F_a \,F_b
K^{a b} V_{1,n} = 0\nonumber\\
{\textstyle{1\over 6}}\e^{a b c d} \,F_b \,F_c \,F_d
V_{1,n} &=& - i F_b\, K^{a b} V_{0,n} +  2i F^a  V_{0,n} - E^a V_{-1,n}
\nonumber\\
F_a\,F_b\, V_{0,n} - {\textstyle{i\over 2}} \e_{a b c d} \, K^{cd}\,
V_{-1,n} &=& 0\,,\qquad F_a \, V_{-1,n} = 0\,;
\end{eqnarray}
\begin{eqnarray}
\bar F^4 V_{n,1} &=& \bar F_{\bar a}\bar F_{\bar b}
\bar K^{\bar a\bar b} V_{n,1} = 0\nonumber\\
{\textstyle{1\over 6}} \e^{\bar a\bar b\bar c\bar d}\bar
F_{\bar b}\bar F_{\bar c} \bar F_{\bar d}
V_{n,1} &=& -i \bar F_{\bar b} \bar K^{\bar a\bar b} V_{n,0}
+ 2i \bar F^{\bar a}  V_{n,0} - \bar E^{\bar a} V_{n,-1}\nonumber\\
\bar F_{\bar a}\bar F_{\bar b} V_{n,0}
- {\textstyle{i\over 2}} \bar\e_{\bar a \bar b \bar c \bar d}
\, \bar K^{\bar c\bar d}\, V_{n,-1} &=& 0\,,
\qquad \bar F_{\bar a} \, V_{n,-1} = 0.\nonumber\\
\end{eqnarray}
There is also a spin zero condition constructed from the Laplacian
\begin{equation}
(\,F_a \, E_a \,+ {\textstyle{1\over 8}}\e_{a b c d} \, K^{ab}\, K^{cd}\,)
\, V_{n,m} =
(\,\bar F_{\bar a} \, \bar E_{\bar a} \,
+ {\textstyle{1\over 8}}\bar\e_{\bar a \bar b \bar c \bar d} \,
\bar K^{\bar a \bar b}\, K^{\bar c\bar d}\,)
\, V_{n,m} = 0\,.
\end{equation}
We derived$^2$ 
the curved space equations (7-9)
by deforming the equations
for the flat case (3-5),
by requiring invariance under the $PSU(2|2)$
transformations (10) that replace the $D=6$ super Poincare transformations
of flat space. The Lie algebra of the supergroup $PSU(2|2)$
contains six even elements $K_{ab}\in SO(4)$ and eight odd ones
$E_a, F_a$.
They generate the infinitesimal symmetry transformations of the
constraint equations:
\begin{eqnarray}&\Delta_a^-\, V_{m,n}=F_a \, V_{m,n}\,,\quad
\Delta_{ab}\, V_{m,n} = K_{ab} \, V_{m,n}&\nonumber\\
&\Delta_a^+\, V_{1,n} = E_a \, V_{1,n}\,,\quad
\Delta_a^+\,V_{0,n}=E_a\,V_{0,n} + i F_a V_{1,n}\,,\quad
\Delta_a^+\, V_{-1,n} = E_a \, V_{-1,n} - i F_a V_{0,n}\,.&\nonumber\\
\end{eqnarray}
We write $E_a$, $F_a$, and $K_{ab}$ for the operators
that represent the left action of $e_a, $ $f_a$, and $t_{ab}$ on
$g$.
In the above coordinates,
\begin{eqnarray}
F_a &=& {d\over d\t^a}\,,\qquad
K_{ab} = -\t_a {d\over d\t^b} + \t_b {d\over d\t^a} + t_{Lab}\nonumber\\
E_a &=& {\textstyle{1\over 2}}\epsilon_{abcd}\,\t^b\,
( t_{L}^{cd} - \t^c {d\over d\t_d}\,) + h_{a\bar b} 
{d\over d\bar\t_{\bar b}}\,,\nonumber\\
\end{eqnarray}
where we have introduced an operator $t_L$
that generates the left action of $SU(2)\times SU(2)$ on $h$ alone,
without acting on the $\theta$'s. Here
\begin{equation}
g = g (x,\t\,, \bar \t)\,
= e^{\t^a f_a}\,
e^{{\scriptsize{1\over 2}}
\sigma^{p cd} x_p t_{cd}}\, e^{\bar\t^{\bar a} e_{\bar a}}
= e^{\t^a f_a}\, h(x) \, e^{\bar\t^{\bar a} e_{\bar a}},
\end{equation}
\begin{equation} 
t_{Lab} \, g \, = \, e^{\t^a f_a}\, (-t_{ab})
\, h(x) \, e^{\bar\t^{\bar a} e_{\bar a}}\,,
\end{equation}
and we found (11) by requiring
$F_ag=f_ag$, $E_ag=e_ag$,
$K_{ab}g=-t_{ab}g$. 
Similar expressions$^2$ hold for the right-acting generators
$\bar K_{\bar a\bar b}$, $\bar E_{\bar a}$, and $\bar F_{\bar a}$.

\section{Supersymmetry Algebras}
\noindent
In flat space, the $D=6$ supersymmetry algebra for the left-movers is 
given by
\begin{eqnarray}
\{ q_a^+, q_c^- \} &=& \half\epsilon_{abcd} P^{cd}\nonumber\\
{[ P_{ab}, P_{cd} ]}  &=& 0 \,\,
= [ P_{ab} , q_c^\pm ] = \{q_a^+,q_b^+\}=
\{q_a^-, q_b^-\}\nonumber\\
\end{eqnarray}
where $P_{ab} \equiv \delta_{ac}\delta_{bd} P^{cd}$ and
\begin{eqnarray}
q_a^- &=& \oint F_a (z)\nonumber\\
q_a^+ &=& \oint ( e^{-\rho -i\sigma} F_a(z) + i E_a(z) )\\
P^{ab} &=& \oint \partial x_m (z) \sigma^{mab}\,.\nonumber\\
\end{eqnarray}
In flat space we have $F_a(z) = p_a(z)$ and
$E_a(z) = \half\e_{abcd} \t^b(z) \partial x_m (z) \sigma^{mcd}$.
We distinguish between the currents and their zero moments
$E_a, F_a$ which together with $P_{ab}$ also generate
the flat space supersymmetry algebra
\begin{eqnarray}
{[ P_{ab}, P_{cd} ]} &=& 0 = [ P_{ab}, F_c ] = [ P_{ab}, E_c ]\,, \nonumber\\
\{ E_a , F_b \} &=& \half \epsilon_{abcd} P^{cd}\,,\qquad
\{ E_a, E_b\} = \{F_a, F_b \} = 0\,.\\
\nonumber\end{eqnarray}
On $AdS_3\times S^3$, the Poincare supersymmetry algebra (17) is replaced by
the $PSU(2|2)$ superalgebra 
\begin{eqnarray}
{[ K_{ab}, K_{cd} ] }&=& \delta_{ac} K_{bd} - \delta_{ad} K_{bc} - 
\delta_{bc} K_{ad} + \delta_{bd} K_{ac}\nonumber\\
{[ K_{ab}, E_c ]} &=&  \delta_{ac} E_b - \delta_{bc} E_a\,,\qquad
{[ K_{ab}, F_c ]} =  \delta_{ac} F_b - \delta_{bc} F_a\,\qquad\nonumber\\
\{E_a, F_b\} &=& \half\e_{abcd} K^{cd}\,,\quad
\{E_a, E_b\} = 0 = \{F_a, F_b\}\,. \nonumber\\
\end{eqnarray}
The generators $q_a^\pm$, 
which generate the $AdS$ tranformations (10), 
still have a form similar to (15) but 
$E_a(z,\bar z), F_a(z,\bar z)$, $K_{ab}(z,\bar z)$
are no longer holomorphic and their
zero moments with respect to $z$ satisfy (18).

\section{String Equations for the $AdS$ Vertex Operator Components}
\noindent
The $AdS$ supersymmetric constraints (7-9) imply$^2$
\begin{equation}
F_a \,F_b \,K^{a b} V_{1,1} = 0\,,\qquad
\bar F_{\bar a}\bar F_{\bar b}\,\bar K^{\bar a\bar b} V_{1,1}
\end{equation}
\begin{equation}
(\, F_a \, E_a \,+ {\textstyle{1\over 8}}\e_{a b c d} \, K^{ab}\, K^{cd}\,)
\, V_{1,1} =
(\,\bar F_{\bar a} \, \bar E_{\bar a} \,
+ {\textstyle{1\over 8}}\bar\e_{\bar a \bar b \bar c \bar d} \,
\bar K^{\bar a \bar b}\, K^{\bar c\bar d}\,)
\, V_{1,1} = 0\,.
\end{equation}
The vertex operators  
$V_{-1,1}, V_{1,-1}, V_{0,-1}, V_{-1,0}, V_{-1,-1}$
can be gauge fixed to zero, and therefore
do not correspond to propagating degrees of freedom.
Furthermore, this gauge symmetry can be used both to set to zero
the components of $V_{1,1}$ with no $\t$'s or no $\bar\t$'s,
and to gauge fix all components of $V_{0,1}, V_{1,0}, V_{0,0}$
that are independent of those of $V_{1,1}$.
The physical degrees of freedom of the massless compactification 
independent vertex operators are thus
described by a superfield
\begin{eqnarray}
V_{1,1}&=&\t^a\bar\t^{\bar a} V^{--}_{a\bar a} +
\t^a\t^b\bar\t^{\bar a}\s^m_{ab} \bar\xi^-_{m\,\bar a}+
\t^a\bar\t^{\bar a} \bar\t^{\bar b}\s^m_{\bar a\bar b}
\xi^-_{m\, a}\nonumber\\
&\hskip15pt +&\t^a\t^b\bar\t^{\bar a} 
\bar\t^{\bar b}\s^m_{ab}\s^n_{\bar a\bar b}
( g_{mn}+b_{mn}+\bar g_{mn}\phi) +
\t^a(\bar\t^3)_{\bar a} A^{-+\,\bar a}_{a}
+(\t^3)_a\bar\t^{\bar a} A^{+-\, a}_{\bar a}\nonumber\\
&\hskip15pt +&\t^a\t^b(\bar\t^3)_{\bar a}\s^m_{ab} \bar\chi_m^{+\,\bar a}+
(\t^3)^a\bar\t^{\bar a} \bar\t^{\bar b}
\s^m_{\bar a\bar b} \chi_m^{+\, a}+
(\t^3)_a(\bar\t^3)_{\bar a} F^{++\, a\bar a}\,.\nonumber\\
\end{eqnarray}
This has the field content of $D=6$, $N=(2,0)$ supergravity
with one supergravity and one tensor multiplet. Further massless multiplets
correspond to the compactification degrees of freedom.  
In flat space,
the surviving constraint equations 
imply that the component fields $\Phi$ are all on shell massless
fields, that is
$\sum_{m=1}^6\p^m \p_m \Phi=0$ as in (5), and in addition
\begin{eqnarray}
\p^m g_{mn} &=& -\p_n\phi\,,\quad  \p^m b_{mn} = 0\,,\quad
\p^m \chi^{\pm b}_m = \p^m \bar\chi^{\pm\bar b}_m = 0\nonumber\\
\partial_{ab} \chi_m^{\pm b} &=& \partial_{\bar a\bar b}
\chi_m^{\pm \bar b} = 0\,,\quad
\partial_{cb} F^{\pm\pm b\bar a} =
\partial_{\bar c\bar b} F^{\pm\pm \bar b a} = 0\,,
\end{eqnarray}
where 
\begin{eqnarray}
F^{+- a\bar a} = \p^{\bar a\bar b}
A_{\bar b}^{+- a}\,,&&
F^{-+ a\bar a} = \p^{a b}
A_b^{-+ \bar a}\,,\quad
F^{-- a\bar a} = \p^{a b} \p^{\bar a\bar b} V_{b\bar b}^{--}\cr
&\chi_m^{-a}& = \p^{ab} \xi_{m b}^-\,, \quad
\bar\chi_m^{- \bar a} = \p^{\bar a\bar b} \bar\xi_{m{\bar b}}^-\,.
\end{eqnarray}
The equations of motion for the flat space vertex operator component fields
describe $D=6$, $N=(2,0)$ supergravity$^4$
expanded around the six-dimensional Minkowski metric.  

In $AdS_3\times S^3$ space corresponding gauge transformations
reduce the number of degrees of freedom in a similar fashion, 
but the Laplacian must be replaced by the $AdS$ Laplacian, and
the constraints are likewise deformed. We focus on the vertex operator 
$V_{11}$ that carries the physical degrees of freedom.
We show the string constraint equations are equivalent to
the $D=6, N=(2,0)$ linearized supergravity equations expanded around
the $AdS_3\times S^3$ metric.      

For the bosonic field components of the vertex operator the $AdS$ constraint
equations result in\pagebreak

\begin{equation}
\Box \,h^g_{\,\,\bar a} \,V^{--}_{ag} =
-4 \,\sigma^m_{ab}\,\sigma^n_{gh}\,\delta^{bh}\, h^g_{\,\,\bar a}\,
G_{mn}
\end{equation}
\begin{equation}
\Box \, h^g_{\,\,\bar a}\,
h^h_{\,\,\bar b}\, \sigma^m_{ab}\,\sigma^n_{gh}\, G_{mn} =
{\textstyle{1\over 4}} \epsilon_{abce} \epsilon_{fghk}
\, \delta^{ch}\,h^f_{\,\,\bar a}\,  h^g_{\,\,\bar b}\,
\, F^{++ e k}
\end{equation}
\begin{equation}
\Box \, h_g^{\,\,\bar a}\,
F^{++ a g} = 0\,,\quad
\Box \, h_g^{\,\,\bar a}\, A_a^{-+  g} = 0\,,\quad
\Box \, h^g_{\,\,\bar a}\, A_g^{+-  a} = 0
\end{equation}
\begin{equation}
\epsilon_{eacd}\, t^{cd}_L\, h^b_{\,\,\bar a}\, A_b^{+-a} = 0\,,\qquad
\epsilon_{\bar e\bar b\bar c\bar d}\, t^{\bar c\bar d}_R\,
h_a^{\,\,\bar a}  \,A_{\bar a}^{-+ \bar b} = 0
\end{equation}
\begin{equation}
\epsilon_{eacd}\, t^{cd}_L\, h_b^{\,\,\bar a}\, F^{++ab} = 0\,,\qquad
\epsilon_{\bar e\bar b\bar c\bar d}\, t^{\bar c\bar d}_R\,
h^a_{\,\,\bar a}  \,F^{++\bar a \bar b} = 0
\end{equation}
\begin{equation}
t^{ab}_L\, h^g_{\,\,\bar a}\,  h^h_{\,\,\bar b}\,\sigma^m_{ab}\,\sigma^n_{gh}\,
G_{mn} =\,0\,,\quad
t^{\bar a\bar b}_R\, h^{\,\,\bar g}_a\,  h^{\,\,\bar h}_{b}\,
\sigma^m_{\bar g\bar h}\,\sigma^n_{\bar a \bar b}\, G_{mn}\,
=\,0\,.
\end{equation}

We have expanded $G_{mn}= g_{mn} + b_{mn} + \bar g_{mn}\phi$.
The $SO(4)$ Laplacian is\break
$\Box\equiv\nobreak{\textstyle{1\over 8}} \epsilon_{abcd} \,t^{ab}_L\, t^{cd}_L
\,=\, {\textstyle{1\over 8}} \epsilon_{\bar a\bar b\bar c\bar d}
\,t^{\bar a\bar b}_R\, t^{\bar c \bar d}_R$.    
In order to compare this with supergravity, we need to reexpress
the above formulas containing the right- and left-invariant vielbeins
$t^{ab}_L, t^{\bar a\bar b}_R$ in terms of covariant derivatives $D_p$ on
the group manifold. So we write
\begin{equation}
{\cal T}_L^{cd} \equiv -\sigma^{p\,cd} \, D_p\,,\qquad
{\cal T}_R^{\bar c\bar d} \equiv \sigma^{p\, \bar c\bar d} \, D_p\,\,.
\end{equation}
Acting on a scalar, ${\cal T}_L=t_L$ and ${\cal T}_R=t_R$, since
both just act geometrically.
But they differ in acting on fields that carry spinor or vector indices.
For example, on spinor indices,

\begin{equation}
t_L^{ab} V_e=
{\cal T}_L^{ab} \, V_e + \half \delta^a_e\, \delta^{bc} V_c
- \half \delta^b_e\, \delta^{ac} V_c\,.
\end{equation}
For $AdS_3\times S^3$  we can write the Riemann tensor
and the metric tensor as 
\begin{eqnarray}
\bar R_{mnp\tau} &=&
{\textstyle{1\over 4}} \, (\, \bar g_{m\tau} \bar R_{np} +
\bar g_{np} \bar R_{m\tau} - \bar g_{n\tau } \bar R_{mp} -
\bar g_{mp} \bar R_{n\tau}\,)\nonumber\\
\bar g_{mn} &=&
{\textstyle{1\over 2}}\, \sigma_m^{ab}\,\sigma_{n\,ab}\,.
\end{eqnarray}
The sigma matrices $\s^{m ab}$ satisfy the algebra 
$\s^{m ab} \s^n_{ac} + \s^{n ab} \s^m_{ac}
= \eta^{mn} \delta^b_c$
in flat space, where $\eta^{mn}$ is the six-dimensional Minkowski metric.
Sigma matrices with lowered indices are defined by
$\s^m_{ab} = \half\e_{abcd} \s^{m cd}$, although for other
quantities indices are raised and lowered with $\delta^{ab}$,
so we distinguish $\s^m_{ab}$ from $\delta_{ac}\,\delta_{bd}\,\s^{m cd}$.
In curved space,  $\eta_{mn}$ is  replaced
by the $AdS_3\times S^3$ metric $\bar g_{mn}$. 

We then find from the string constraints that the six-dimensional
metric field $g_{rs}$,
the dilaton $\phi$, and the two-form $b_{rs}$ satisfy 
\begin{eqnarray}
{\textstyle{1\over 2}} D^p D_p b_{rs} &=&
-{\textstyle{1\over 2}} (\sigma_r\sigma^p\sigma^q)_{ab}\delta^{ab}
\,D_p\,
[\,g_{qs} + \bar g_{qs}\phi\,]
+{\textstyle{1\over 2}} (\sigma_s\sigma^p\sigma^q)_{ab}\delta^{ab}
D_p\,
[\,g_{qr} + \bar g_{qr}\phi\,]\nonumber\\
&-&\bar R_{\tau r s \lambda} \, b^{\tau\lambda}\,
-{\textstyle{1\over 2}} \bar R_r^{\,\,\tau}\, b_{\tau s}
-{\textstyle{1\over 2}} \bar R_s^{\,\,\tau}\, b_{r\tau}\nonumber\\
&+& {\textstyle{1\over 4}} F^{++gh}_{\rm asy}
\,\sigma_r^{ab}\sigma_s^{ef}\, \delta_{ah}\delta_{be}\delta_{gf}
\end{eqnarray}
\begin{eqnarray}
{\textstyle{1\over 2}} D^p D_p \,
(\, g_{rs} + \bar g_{rs} \phi \,)\,&=&
-{\textstyle{1\over 2}} (\sigma_r\sigma^p\sigma^q)_{ab}\delta^{ab}
\,D_p b_{qs} \,
+ {\textstyle{1\over 2}} (\sigma_s\sigma^p\sigma^q)_{ab}\delta^{ab}
\,D_p b_{rq} \nonumber\\
&-& \bar R_{\tau r s \lambda} \,
(\, g^{\tau\lambda} + \bar g^{\tau\lambda}\phi \,)\,
-{\textstyle{1\over 2}} \bar R_r^{\,\,\tau}\,
(\, g_{\tau s} + \bar g_{\tau s} \phi \,)
-{\textstyle{1\over 2}} \bar R_s^{\,\,\tau}\,
(\, g_{r\tau } + \bar g_{r\tau }\phi \,)\nonumber\\
&+& {\textstyle{1\over 4}} F^{++gh}_{\rm sym} \,\sigma_{rga}\sigma_{shb}\,
\delta^{ab}\,.\nonumber\\
\end{eqnarray}
This is the curved space version of the flat space zero Laplacian condition
$\p^p \p_p b_{rs} = \p^p \p_p g_{rs} = \p^p \p_p \phi = 0$. 

Four self-dual tensor and scalar pairs come from the string bispinor
fields\break
$ F^{++ ab}, \,V^{--}_{ab}, \,,A^{+- a}_b, \,A^{-+ b}_a$.  
From the string constraint equations they satisfy
\begin{equation}
\sigma^p_{da}\, D_p \, F^{++ab}_{\rm asy} = 0
\end{equation}
\begin{equation} 
{\textstyle{1\over 4}}\,[\, \delta^{Ba} \,\sigma^r_{ga}\,D_r\,
F^{++gH}_{\rm sym} \, - \,
\delta^{Ha} \,\sigma^r_{ga}\,D_r\, F^{++gB}_{\rm sym} \, ]
= -{\textstyle{1\over 4}}\,\, \epsilon^{BH}_{\qquad cd}\,\,
F^{++cd}_{\rm asy}
\end{equation}
We also find
\begin{eqnarray}
&&{\textstyle{1\over 2}}\, D^p D_p\, V_{cd}^{--}
\, -{\textstyle{1\over 2}} \delta^{gh}\sigma^p_{ch}\, D_p \, V_{gd}^{--}
\, +{\textstyle{1\over 2}} \delta^{gh}\sigma^p_{dh}\, D_p \, V_{cg}^{--}
\, +{\textstyle{1\over 4}} \epsilon_{cd}^{\,\,\,\,gh}\, V_{gh}^{--}\nonumber\\
&=&\,-4\,\sigma^m_{ce}\,\sigma^n_{df}\,\delta^{ef}\, G_{mn}\,.\nonumber\\
\end{eqnarray}
The last constraints can be written as 
\begin{eqnarray}
&\epsilon_{eacd}&
t^{cd}_L\, h_b^{\,\,\bar a}\, F^{+-ab} = 0\,\qquad
\epsilon_{\bar e\bar b\bar c\bar d}\, t^{\bar c\bar d}_R\,
h^a_{\,\,\bar a}  \,F^{+-\bar a \bar b} = 0\nonumber\\
&\epsilon_{eacd}& t^{cd}_L\, h_b^{\,\,\bar a}\, F^{-+ab} = 0\,,
\qquad\epsilon_{\bar e\bar b\bar c\bar d}\, t^{\bar c\bar d}_R\,
h^a_{\,\,\bar a}  \,F^{-+\bar a \bar b} = 0\nonumber\\
\end{eqnarray}
where
\begin{eqnarray}
&F^{+-a\bar a}&\equiv \delta^{\bar a\bar b} \,
A_{\bar b}^{+- a} + t_R^{\bar a\bar b} \, A_{\bar b}^{+- a}\nonumber\\
&F^{-+a\bar a}&\equiv \delta^{ab} \,
A_{b}^{-+ \bar a} + t_L^{ab} \, A_b^{-+ \bar a}\,,\nonumber\\
\end{eqnarray}
so $F^{+-ab}$ and $F^{-+ab}$
satisfy equations similar to those for  $F^{++ab}$.  

Independent conditions on the fermion fields are
\begin{eqnarray}
&&\Box \, h_a^{\,\,\bar g}\,
\s^m_{\bar a\bar b}\, \xi^-_{m \bar g}
= - \s^m_{\bar g\bar h}\,\e_{\bar e\bar d\bar a\bar b} \,
h_a^{\,\,\bar h}\, \delta^{\bar g\bar d} \,\bar\chi_m^{+\bar e}\nonumber\\
&&\Box \, h^g_{\,\,\bar a}\,
\s^m_{a b}\, \bar\xi^-_{m g}
= - \s^m_{gh}\,\e_{edab} \,
h^h_{\,\,\bar a}\, \delta^{gd} \,\chi_m^{+ e}\nonumber\\
&&t_L^{\,\, ab} h^g_{\,\bar a}\, \sigma^m_{ab}\,
\bar\xi_{m g}^{-} \, = 0\,,\qquad
t_R^{\,\, \bar a\bar b}\,  h_a^{\,\bar g}\,\sigma^m_{\bar a\bar b}\,
\xi_{m \bar g}^{-} \, = 0\nonumber\\
&&t_L^{\,\, ab} \sigma^m_{ab}\, h_g^{\,\bar a}\,
\bar\chi_m^{+\,g} \, = 0\,,\qquad
t_R^{\,\, \bar a\bar b}\, \sigma^m_{\bar a\bar b}\,
h^a_{\,\bar g}\, \chi_m^{+\,\bar g} \, = 0\nonumber\\
&&\epsilon_{deab} t_L^{\,ab}\,
h^g_{\,\,\bar a}\, h^h_{\,\,\bar b}\,\sigma^m_{gh}\,
\chi_m^{+\,e} \, = 0\,,\qquad
\epsilon_{\bar d\bar e\bar a\bar b}\, t_R^{\,\bar a\bar b}\,
h_a^{\,\,\bar g}\, h_b^{\,\,\bar h}\,\sigma^m_{\bar g\bar h}
\bar\chi_m^{+\,\bar e} \, = 0\,.\nonumber\\
\end{eqnarray}

\section{Comparison With Linearized $AdS$ Supergravity Equations}
We now show that the  $AdS_3\times S^3$
supersymmetric vertex operator constraint equations  
are equivalent to the linearized supergravity equations
for the supergravity multiplet and one tensor multiplet
of $D=6$, $N=(2,0)$ supergravity$^4$ expanded around the
$AdS_3\times S^3$ metric and a self-dual three-form.
We give the identification of the string vertex operator components
in terms of the supergravity fields. 

We will see that the two-form $b_{mn}$ is a linear combination of
{\it all} the
oscillations corresponding to the five self-dual tensor
fields and the anti-self-dual tensor field, {\it including
the oscillation with non-vanishing background.}
In flat space, $b_{mn}$ corresponds to a state in the Neveu-Schwarz sector.
In our curved space case,
the string model describes vertex operators for $AdS_3$ background
with Ramond-Ramond flux. When matching the vertex operator component fields
with the supergravity oscillations, we find that not only the
bispinor $V^{--}_{ab}$ (which is a Ramond-Ramond field in the
flat space case),  but also the tensor $b_{mn}$
include supergravity oscillations with non-vanishing self-dual background. 

The linearized supergravity equations are given by
\begin{equation}D^p D_p \phi^i ={\textstyle{2\over 3}}\,
\bar H^i_{prs}\, g^{6prs}
\end{equation}
\begin{eqnarray}
&\half D^p D_p & h_{rs}
- \bar R_{\tau r s \lambda} \, h^{\tau\lambda}
+ \half \bar R_r^\tau\, h_{\tau s}
+ \half \bar R_s^\tau\, h_{\tau r}
- \half D_s D^p h_{pr}
- \half D_s D^p h_{pr} + \half D_rD_s h^p_p\nonumber\\
&=&
\,- \bar H^{i\,\,\,pq}_r\, g^i_{spq}\,
- \bar H^{i\,\,\,pq}_s\, g^i_{rpq}\,
+ 2 h^{pt}\,\bar H^{i\,\,\,q}_{rp}\,\bar H^i_{stq}\nonumber\\
\end{eqnarray}
\begin{eqnarray}
D^p H_{prs}
&=& -2  \,\bar H^i_{prs}\, D^p\phi^i\nonumber\\
&\hskip10pt +& B^i\,  [ - \bar H^{i\,\,pq}_r D_p\, h_{qs}
+ \bar H^{i\,\,pq}_s D_p\, h_{qr}
+ \bar H^{i\,\,\,q}_{rs} D^p h_{pq}
-{\textstyle{1\over 2}}
\bar H^{i\,\,\,q}_{rs} D_q h^p_{\hskip5pt p}]\nonumber\\
\end{eqnarray}
where we have defined
$H_{prs} \equiv  \,g^6_{prs} + B^i \,g^i_{prs}$
as a combination of the supergravity exact forms $g^6\equiv db^6,
g^i\equiv d b^i$, since we will equate this with the string
field strength $H=db$. We will choose $B^1 = 2$. 
In zeroeth order, the equations are
$\bar R_{rs} = - \bar H^i_{rpq}\,\bar H^{i\,\,pq}_s$. 

We define the vertex operator components
in terms of the supergravity fields 
$g^i_{prs}, g^6_{prs}, h_{rs}, \phi^i$ , $1\le i\le 5$, (and $2\le I\le 5$)
as
$$H_{prs}\equiv  \,g^6_{prs} + 2 \,g^1_{prs}\,+ B^I \,g^I_{prs}$$
$$g_{rs} \equiv \,h_{rs}
-{\textstyle{1\over 6}}\bar g_{rs}\, h^\lambda_{\,\,\lambda}$$
$$\phi = -{\textstyle{1\over 3}}\, h^\lambda_{\,\,\lambda}$$
\begin{eqnarray}
F^{++ab}_{\rm sym} &= &\,\, {\textstyle{2\over 3}}
(\sigma_p\sigma_r\sigma_s)^{ab}
\, B^I \,\,g^I_{prs} +  \delta^{ab}\,\phi^{++}\nonumber\\
F^{++ab}_{\rm asy} &=&\,\,\sigma^{p\,ab} \, D_p\,\phi^{++}\nonumber\\
\phi^{++} &=&  4 C^I\,\,\phi^I\nonumber\\
\end{eqnarray}
which follows from choosing the graviton trace
$h^\lambda_{\hskip3pt\lambda}$ to satisfy $\phi^1 -
h^\lambda_{\,\,\lambda} \,\equiv\, - 2\, C^I\phi^I$,
and we have used $H_{prs} \equiv \partial_p b_{rs} +  \partial_r b_{sp} +
\partial_s b_{pr}$. 

The combinations $C^I\phi^I$ and $B^I g^I_{prs}$ reflect the
$SO(4)_{\rm R}$ symmetry of the $D=6, N=(2,0)$ theory on $AdS_3\times S^3$.
We relabel $C^I = C^I_{++}$, $B^I = B^I_{++}$.
To define the remaining string components in terms of supergravity fields,
we consider linearly independent quantities $C_\ell^I\phi^I$,
$B_\ell^I  g^I_{prs}$, $\ell = ++,+-, -+, --$.
\begin{eqnarray}
F^{+-ab}_{\rm sym} &= &\,\, {\textstyle{2\over 3}}
(\sigma_p\sigma_r\sigma_s)^{ab}
\, B_{+-}^I \,\,g^I_{prs} +  \delta^{ab}\,\phi^{+-}\nonumber\\
F^{+-ab}_{\rm asy} &= &\,\,\sigma^{p\,ab} \, D_p\,\phi^{+-}\nonumber\\
\phi^{+-} &= & 4 C_{+-}^I\,\,\phi^I\nonumber\\
F^{-+ab}_{\rm sym} &=  &\,\, {\textstyle{2\over 3}}
(\sigma_p\sigma_r\sigma_s)^{ab}
\, B_{-+}^I \,\,g^I_{prs} +  \delta^{ab}\,\phi^{-+}\nonumber\\  
F^{-+ab}_{\rm asy} &= &\,\,\sigma^{p\,ab} \, D_p\,\phi^{-+}\cr
\phi^{-+} &=& 4 C_{-+}^I\,\,\phi^I\, .\nonumber\\
\end{eqnarray}
$V^{--}_{ab}$ is given in terms of the fourth tensor/scalar pair
$C_{--}^I\,\,\phi^I$, $B_{--}^I g_{mnp}^I$ through 
\begin{equation}
D^p D_p\, V_{cd}^{--}
\, - \delta^{gh}\sigma^p_{ch}\, D_p \, V_{gd}^{--}
\, + \delta^{gh}\sigma^p_{dh}\, D_p \, V_{cg}^{--}
\, +{\textstyle{1\over 2}} \epsilon_{cd}^{\,\,\,\,gh}\, V_{gh}^{--}
=\,-8\,\sigma^m_{ce}\,\sigma^n_{df}\,\delta^{ef}\, G_{mn}\,.
\end{equation}
These field definitions allow us to identify the string constraint equations 
for the $AdS_3$ vertex operators as precisely those which require 
the vertex operator field components to satisfy the 
linearized supergravity equations reviewed 
in this section. 

Likewise, the fermion constraints 
imply the linearized AdS supergravity equations for the gravitinos
and spinors, due to the above correspondence for the bosons and the
supersymmetry of the two theories.

\nonumsection{Acknowledgements}
\noindent
This was supported in part by the  U.S. Department of Energy,
Grant No. DE-FG 05-85ER40219/Task A. It reviews work done in collaboration with
E. Witten.

\nonumsection{References}
\noindent

\end{document}